\documentclass[prx,twocolumn,superscriptaddress,showpacs,floatfix,longbibliography]{revtex4-1}
\usepackage{graphicx}
\usepackage{bm}
\usepackage{color,soul}
\usepackage{ulem}
\usepackage{amsmath,amssymb,amsfonts,float,graphics,epsfig,epstopdf,color,verbatim,tabularx,bm,multirow,appendix,hyperref}
\begin{document}

%\preprint{}
\title{Evidences for the topological order in a kagome antiferromagnet}
\author{Yuan Wei}
\affiliation{Beijing National Laboratory for Condensed Matter Physics, Institute of Physics, Chinese Academy of Sciences, Beijing 100190, China}
\affiliation{School of Physical Sciences, University of Chinese Academy of Sciences, Beijing 100190, China}
\author{Zili Feng}
\affiliation{Beijing National Laboratory for Condensed Matter Physics, Institute of Physics, Chinese Academy of Sciences, Beijing 100190, China}
\affiliation{School of Physical Sciences, University of Chinese Academy of Sciences, Beijing 100190, China}
\author{Wiebke Lohstroh}
\affiliation{Heinz Maier-Leibnitz Zentrum (MLZ), Technische Universit$\ddot{a}$t M$\ddot{u}$nchen, Garching D-85747, Germany}
\author{D. H. Yu}
\affiliation{Australian Nuclear Science and Technology Organization, Lucas Heights, New South Wales 2232, Australia}
\author{Duc Le}
\affiliation{ISIS Facility, Rutherford Appleton Laboratory, Chilton, Didcot OX11 0QX, United Kingdom}
\author{Clarina dela Cruz}
\affiliation{Neutron Scattering Division, Neutron Sciences Directorate, Oak Ridge National Laboratory, Oak Ridge, Tennessee 37831, USA}
\author{Wei Yi}
\affiliation{International Center for Materials Nanoarchitectonics, National Institute for Materials Science, 1-1 Namiki, Tsukuba, Ibaraki 305-0044, Japan}
\author{Z. F. Ding}
\affiliation{State Key Laboratory of Surface Physics, Department of Physics, Fudan University, Shanghai 200433, China}
\author{J. Zhang}
\affiliation{State Key Laboratory of Surface Physics, Department of Physics, Fudan University, Shanghai 200433, China}
\author{Cheng Tan}
\affiliation{State Key Laboratory of Surface Physics, Department of Physics, Fudan University, Shanghai 200433, China}
\author{Lei Shu}
\affiliation{State Key Laboratory of Surface Physics, Department of Physics, Fudan University, Shanghai 200433, China}
\affiliation{Collaborative Innovation Center of Advanced Microstructures, Nanjing 210093, China}
\author{Yan-Cheng Wang}
\affiliation{School of Physical Science and Technology, China University of Mining and Technology, Xuzhou Jiangsu, 221116, People’s Republic of China}
\author{Han-Qing Wu}
\affiliation{School of Physics, Sun Yat-Sen University, Guangzhou 510275, China}
\author{Jianlin Luo}
\affiliation{Beijing National Laboratory for Condensed Matter Physics, Institute of Physics, Chinese Academy of Sciences, Beijing 100190, China}
\affiliation{School of Physical Sciences, University of Chinese Academy of Sciences, Beijing 100190, China}
\affiliation{Songshan Lake Materials Laboratory , Dongguan, Guangdong 523808, China}
\author{Jia-Wei Mei}
\affiliation{Shenzhen Institute for Quantum Science and Engineering, and Department of Physics, Southern University of Science and Technology, Shenzhen 518055, China}
\author{Fang Yang}
\affiliation{School of Physics, Key Laboratory of Micro-Nano Measurement-Manipulation and Physics (Ministry of Education), Beihang University, Beijing 100191, China}
\author{Xian-Lei Sheng}
\affiliation{School of Physics, Key Laboratory of Micro-Nano Measurement-Manipulation and Physics (Ministry of Education), Beihang University, Beijing 100191, China}
\author{Wei Li}
\affiliation{School of Physics, Key Laboratory of Micro-Nano Measurement-Manipulation and Physics (Ministry of Education), Beihang University, Beijing 100191, China}
\affiliation{International Research Institute of Multidisciplinary Science, Beihang University, Beijing 100191, China}
\author{Yang Qi}
\affiliation{Center for Field Theory and Particle Physics,
	Department of Physics, Fudan University, Shanghai 200433, China}
\affiliation{Collaborative Innovation Center of Advanced Microstructures, Nanjing 210093, China}
\author{Zi Yang Meng}
\email{zymeng@iphy.ac.cn}
\affiliation{Beijing National Laboratory for Condensed Matter Physics, Institute of Physics, Chinese Academy of Sciences, Beijing 100190, China}
\affiliation{Department of Physics and HKU-UCAS Joint Institute of Theoretical and Computational Physics, The University of Hong Kong, Pokfulam Road, Hong Kong, China}
\affiliation{Songshan Lake Materials Laboratory , Dongguan, Guangdong 523808, China}
\author{Youguo Shi}
\email{ygshi@iphy.ac.cn}
\affiliation{Beijing National Laboratory for Condensed Matter Physics, Institute of Physics, Chinese Academy of Sciences, Beijing 100190, China}
\affiliation{Center of Materials Science and Optoelectronics Engineering, University of Chinese Academy of Sciences, Beijing 100049, China}
\affiliation{Songshan Lake Materials Laboratory , Dongguan, Guangdong 523808, China}
\author{Shiliang Li}
\email{slli@iphy.ac.cn}
\affiliation{Beijing National Laboratory for Condensed Matter Physics, Institute of Physics, Chinese Academy of Sciences, Beijing 100190, China}
\affiliation{School of Physical Sciences, University of Chinese Academy of Sciences, Beijing 100190, China}
\affiliation{Songshan Lake Materials Laboratory , Dongguan, Guangdong 523808, China}

\begin{abstract}
A $Z_2$ quantum spin liquid hosts one of the simplest topological orders and exhibits many exotic properties due to long-range quantum entanglements. Its elementary excitations are anyons such as spinons carrying fractionalized spin quantum number and visons carrying emergent $Z_2$ gauge flux. However, experimental detection of these anyons remains elusive. The difficulties lie not only in the fact that there exists few candidates for $Z_2$ quantum spin liquids but also in that visons are magnetically inert hence immune to available experimental techniques. Here we have studied the spin excitations and specific heats of kagome-lattice antiferromagnet Cu$_4$(OH)$_6$FBr and Cu$_3$Zn(OH)$_6$FBr, which consists of two-dimensional Cu$^{2+}$ kagome layers with either Cu$^{2+}$ or Zn$^{2+}$ ions in between. By combining the first principle calculations and inelastic neutron scattering data in the former, we show that the dominate couplings in magnetically ordered Cu$_4$(OH)$_6$FBr are between the nearest neighbor spins within the kagome planes, and the kagome and interlayer spin systems are essentially decoupled above the antiferromagnetic transition temperature. The intrinsic spin excitations and specific heats of the kagome layers for Cu$_3$Zn(OH)$_6$FBr are thus derived by removing the contributions from the residual interlayer Cu$^{2+}$ magnetic impurities.  Accordingly, the kagome spin system exhibits spin continuum with momentum-dependent spin gap and a large magnetic entropy at low temperature that is insensitive to magnetic field, which can be understood as the evidences of spinons and visons in this kagome quantum spin liquid candidate. Our results suggest the existence of the $Z_2$ anyons in the material, and therefore provide a comprehensive set of evidences for the $Z_2$ topological order in kagome quantum spin liquid and bring their choreographed entanglement dances to the stage of real materials.
	
\end{abstract}


%\pacs{}

\maketitle

\section{introduction}

Topological insulators and superconductors have attracted great attentions in condensed matter physics and material sciences ~\cite{HasanMZ10,QiXL11}. These states are associated with classical topology and denoted as the symmetry protected topological states. There is another class of states, dubbed topological orders, which has also been intensely studied in theory \cite{WenXG17,SavaryL17,ZhouY17,WenXG19}. It is referred to as quantum topology, since it results from long-range quantum entanglement and features fractionalized excitations, which are called anyons because they carry fractionalized statistics \cite{WenXG19}.
This exotic class of matter does not require the labeling of order parameter and spontaneous symmetry breaking so they are beyond Landau's paradigm of phases and phase transitions.
Moreover, their topological nature does not need to be protected by any global symmetry, as they are robust against any local perturbation.
Therefore, they are seriously expected to be the building block for topological quantum computation~\cite{KitaevAY03,NayakC08,RaussendorfR07,KouSP09}.
Although not protected by any symmetry, a topological order can be further enriched by symmetries.
In particular, the fractionalized excitations can carry fractionalized quantum numbers of global symmetries \cite{Cheng2016,SunGY18}.

Up to now, only quantum Hall states can be unambiguously identified as topological orders with chiral quantum topology \cite{WenXG90,WenXG90b} . 
In principle, the topological orders can also be found in quantum spin liquids (QSL), which are conceptually connected to the theoretical understanding of high-$T_c$ superconductivity and, as mentioned above, the material realization of topological quantum computing \cite{ZhouY17,WenXG19,RaussendorfR07,KouSP09,ReadN91,WenXG91,SachdevS92,KitaevAY03}. Among various QSLs, a $Z_2$ QSL is probably the simplest one, as all its excitations are gapped and the low-lying anyonic quasiparticles therein have relatively simple mathematical descriptions. However, despite the transparent theoretical understanding, there are few candidate materials that may host a $Z_2$ QSL \cite{SavaryL17,NormanMR16,VriesMA09,HanTH12,HanTH16,FuM15,FengZL17,BroholmC20,ZiliFeng2019,JJWen2019,WeiYuan20Magnetic,WeiYuan20Nonlocal}. One of the reasons behind the difficulty of finding more candidates is because their properties, or in modern language, the choreographed entanglement dances of anyons carrying fractionalized quantum number that are dedicated by the emergent gauge structures \cite{WenXG19}, are not subjected to direct experimental probes which usually couple to the local order parameters related with symmetry breaking such as spontaneous magnetization.

A two-dimensional (2D) kagome lattice antiferromagnet is expected to host a $Z_2$ QSL and thus the $Z_2$ topological order due to strong 
geometrical and quantum frustrations \cite{SachdevS92,JiangHC08,YanS11,JiangHC12,DepenbrockS12,MeiJW17,JiangS16,LiaoHJ17,HeYC17,JJWen2019}. In this case, the ultimate experimental discovery of a $Z_2$ QSL relies on the signature of fractionalized excitations -- spinons (carrying $S$ = 1/2 quantum number) and visons (carrying emergent $Z_2$ guage flux with $S$ = 0), and the dynamical coupling effects among these anyons mediated by the emergent $Z_2$ gauge field. 
Theoretically, these effects have been hypothesized from many perspectives~\cite{ReadN91,WenXG91,KitaevAY03,SenthilT00,NikolicP03,HuhY11,PunkM14,SunGY18,Becker2018,YCWang2018,YCWang2020Vestigial}, such as their translational symmetry fractionalization~\cite{SunGY18,Becker2018} or various condensation channels~\cite{YCWang2020Vestigial}, but the unfortunate reality is that their direct experimental detection is still no where to find. Experimentally, herbertsmithite ZnCu$_3$(OH)$_6$Cl$_2$ is probably the most promising candidate for a QSL, but it is still not conclusive whether its excitations are gapped or not \cite{VriesMA09,HanTH12,HanTH16,FuM15,Khuntia2019}. Because of the influence of magnetic impurities, even for single-crystal herbertsmithite, the spin excitations look gapless at the first place~\cite{HanTH12}. Later, it is from nuclear magnetic resonance (NMR) experiment that the spin excitations are gapped~\cite{FuM15}. However, since the analysis is rather complicated due to the presence of five peaks of $^{17}$O in the NMR spectra, it has also been suggested that the ground state of the herbertsmithite is gapless \cite{Khuntia2019}. Detailed analysis on the inelastic neutron scattering (INS)data also shows a gap at $(1,0,0)$ by subtracting the intensities at $(1,0,0)$ from those at $(2,0,0)$ times a parameter to make sure the low-energy intensities are close to zero \cite{HanTH16}, but the choice of the parameter is still not objective enough. 

In this work, we try to clarify the confusion in the field by providing a tentative solution to this problem, i.e., we find experimental evidences of a $Z_2$ topological order or QSL in a 2D kagome material Cu$_3$Zn(OH)$_6$FBr. Both its crystal structure and magnetic properties are very similar to those of herbertsmithite \cite{FengZL17}, such as very large exchange coupling ($\sim$200 K) and no magnetic order down to 50 mK. It consists of 2D Cu$^{2+}$ kagome layers with Zn$^{2+}$ ions between the layers as shown in Fig. \ref{fig:spinexchn}(a). Like in herbertsmithite, there is always a few percent of Cu$^{2+}$ ions sitting on the Zn$^{2+}$ sites, acting as magnetic impurities. When all Zn$^{2+}$ ions are replaced by Cu$^{2+}$ ions, the material is a natural mineral dubbed barlowite and shows slight lattice distortion below room temperature and an antiferromagnetic order at $T_N \sim$ 15 K \cite{HanTH14,FengZL18,SmahaRW20}.
Previous NMR measurements on Cu$_3$Zn(OH)$_6$FBr suggest a spinon gap about 0.65 meV, whose magnetic field dependence is consistent with fractionalized $S$ = 1/2 spinon excitations \cite{FengZL17,XGWen2017}. The measurements were done on $^{19}$F, which shows only one peak and is not sensitive to magnetic impurities due to its position. 

Here we further studied the spin excitations and specific heats of Cu$_3$Zn(OH)$_6$FBr to reveal its ground states. One thing we have to overcome is the influences from the residual interlayer Cu$^{2+}$ spins as magnetic impurities, which contribute to the low-energy spin excitations and low-temeprature speific heats as for the case in herbertsmithite \cite{NormanMR16}. To do so, we have compared first principle calculations and INS data in magnetically ordered Cu$_4$(OH)$_6$FBr to show that the couplings between the interlayer and kagome spins are one order of magnitude smaller than those within the kagome planes. Moreover, the spin excitations from these two spin systems become separated from each other above $T_N$, which is further confirmed in paramagnetic Cu$_3$Zn(OH)$_6$FBr with a few percent of residual interlayer Cu$^{2+}$. Accordingly, we are able to compare two Cu$_3$Zn(OH)$_6$FBr samples with different concentrations of magnetic impurities and simply subtract their contributions. 

Our results show that the spin excitations ($S$ = 1) of Cu$_3$Zn(OH)$_6$FBr are  gapped with a gap value at $Q$ = 0 twice of the spinon gap measured in NMR. Moreover, the low-temperature specific heat from the kagome magnetic system shows no magnetic field dependence, which suggests it is dominated by the $S$ = 0 excitations -- consistent with the existence of vison excitations which carry zero spin quantum number with a smaller gap compared with that of the spinon. These results reveal the presence of gapped spinon and vison excitations as well as the dynamical coupling between these anyonic excitations, and set up a comprehensive theoretical and experimental protocol to measure materials such as Cu$_3$Zn(OH)$_6$FBr and to provide evidence for a Z$_2$ topological order therein. 

\section{experiments}

Polycrystalline Cu$_4$(OD)$_6$FBr and Cu$_3$Zn(OD)$_6$FBr were synthesized by the hydrothermal method with a mixture of Cu$_2$(OD)$_2$CO$_3$, ZnF$_2$, ZnBr$_2$ and CuBr$_2$ sealed in a reaction vessel with heavy water. From now on, we will use Cu4 and Cu3Zn to refer Cu$_4$(OD)$_6$FBr and Cu$_3$Zn(OD)$_6$FBr, respectively. For Cu3Zn, two sets of samples labeled with S1 and S2 were grown at 230 and 200 $^\circ$C, respectively. The inductively coupled plasma (ICP) measurements suggest that the Zn contents in S1 and S2 are 0.89 $\pm$ 0.04 and 1.10 $\pm$ 0.07, respectively, where error bars are from the minimum and maximum concentrations measured for several samples. The heat capacity was measured by a Physical Properties Measurement System (PPMS, Quantum Design). Neutron diffraction pattern was measured on the HB-2A diffractometer at HFIR, USA with the wavelength of 2.4103 \AA. The INS experiments were carried out on the TOFTOF spectrometer at FRM-II, Germany, the PELICAN spectrometer at ANSTO, Australia and the MARI at ISIS, the United Kingdom. The absolute values of the spin excitations for the Cu3Zn-S2 sample measured at TOFTOF are obtained by normalizing to a vanadium standard sample. The intensities of spin excitations measured at other instruments are further normalized by comparing the integration of Bragg peaks with that on TOFTOF \cite{supp}.  In representing the neutron data, the intensity $I(Q,\omega)$ is the raw data including both the signal from the sample and the backgrounds. On the other hand, the neutron scattering function $S(Q,\omega)$ is obtained by taking advantage of the neutron-gain-energy process and only contains the excitations from the sample without backgrounds \cite{supp}. Moreover, the value of $S(Q,\omega)$ has been normalized by the vanadium sample \cite{supp}.

\section{results}

\subsection{Spin couplings in compounds Cu$_3$Zn(OH)$_6$FBr and Cu$_4$(OH)$_6$FBr}

\begin{figure}[tbp]
\includegraphics[width=\columnwidth]{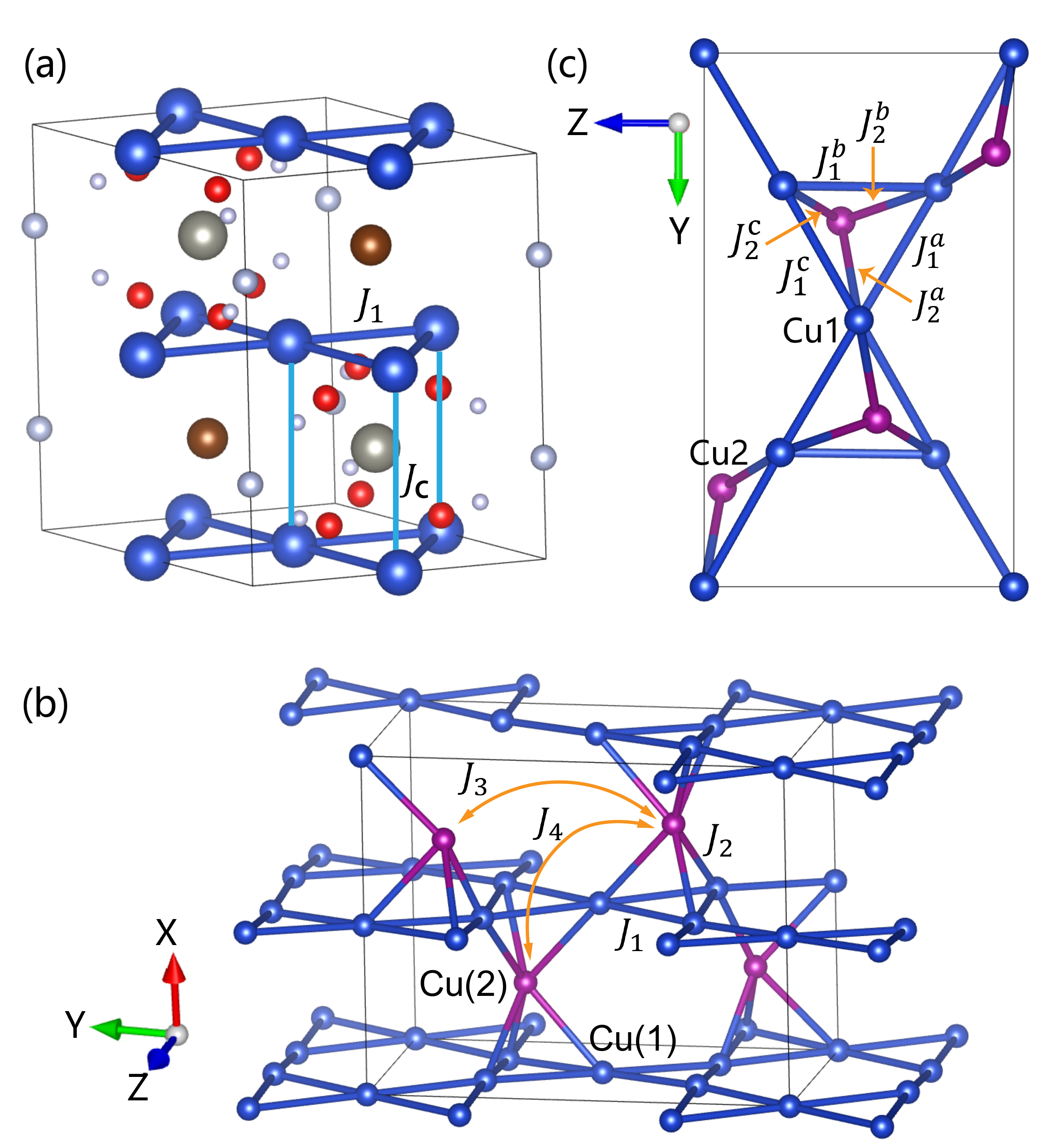}
 \caption{(a) The lattice structure of Cu3Zn, 
 with the spin couplings denoted as $J_1$ 
 (within kagome plane) and $J_c$ (inter-plane).
 (b) Lattice structure and couplings 
 ($J_1$, $J_2$, $J_3$ and $J_4$) in the compound Cu4. 
 Cu(1) represents the kagome Cu$^{2+}$ ions and Cu(2) 
 the interlayer ones. (c) Due to the lattice 
 distortion, in Cu4 structure the couplings $J_1$ 
 and $J_2$ can have different strengthens on the
 three non-equivalent bonds. }
\label{fig:spinexchn}
\end{figure}

To begin with, we show the first principle calculation results, which
suggest that the kagome and interlayer spin systems are only weakly coupled.
In order to estimate the spin exchange coupling strength between Cu$^{2+}$ ions in the two compounds Cu3Zn and Cu4, we perform DFT+U calculations \cite{Anisimov1991,LDAU,dudarev1998} 
and compare the free energies of different magnetic structures. 
Our calculations used the PBE \cite{PBE1996} functional and 
the $U_{\rm eff}$ was set to 5 eV. Figure \ref{fig:spinexchn} shows the 
definitions of different superexchange couplings in the lattice. 
Considering that the Cu$^{2+}$ ions in Cu3Zn constitute 
a standard kagome lattice plane with perfect $C_3$ 
symmetry, it is reasonable to assume the couplings 
are equal on equivalent kagome bonds, 
as shown in Fig. \ref{fig:spinexchn}(a). By comparing different classical 
spin configurations, including the ferromagnetic, 
ferrimagnetic (up-up-down), and co-planer antiferromagnetic ones,
we estimate the Heisenberg coupling strength $J_1$ between 
two nearest-neighbouring (NN) Cu$^{2+}$ ions in 
the same kagome plane as about 7 meV, 
and the interplane coupling $J_c$ as about 0.6 meV. 
As $J_c$ is more than one order of magnitude smaller than $J_1$, 
our DFT calculations thus confirm the excellent two-dimensionality 
of the frustrated kagome magnet Cu3Zn.

\begin{table}[bp]
\caption{\label{tab:table1}
Bond lengths and spin couplings in the compound Cu4, 
see Fig.~\ref{fig:spinexchn} (b,c) for the illustration of lattice structure and couplings.
}
\begin{ruledtabular}
\begin{tabular}{lcr}
\textrm{Coupling}&
\textrm{$d_{Cu-Cu}$(\AA)}&
\textrm{Value(meV)}\\
\colrule
$J_1^a$	& 3.324	& 11.588\\
$J_1^b$	& 3.345	& 13.386\\
$J_1^c$	& 3.334	& 4.554\\
\colrule
$J_2^a$	& 3.173	& -0.444\\
$J_2^b$	& 3.133	& -1.476\\
$J_2^c$	& 2.824	& -0.586\\
\colrule
$J_3$ & 6.301 &	0.013\\
$J_4$ &	6.346 & -0.042\\ 
\end{tabular}
\end{ruledtabular}
\label{coupling}
\end{table}

On the other hand, in the compound Cu4 there exists a lattice distortion 
at low temperature, which breaks the $C_3$ symmetry of the kagome lattice. 
Therefore, we employ a more sophisticated four-spin-states method \cite{Xiang2011} 
to better evaluate the coupling parameters [c.f., Fig. \ref{fig:spinexchn}(b) and \ref{fig:spinexchn}(c)] and also list the  
results in Tab.~\ref{coupling}. Our DFT results show that the slight lattice 
distortion can actually lead to a significant anisotropy in $J_1$ and $J_2$ couplings.
In particular, $J_2^b$ is clearly distinct from other two $J_2$ couplings,
all of which are of FM type. Moreover, we find that 
$J_3$ and $J_4$ couplings between interlayer Cu(2) 
are significantly smaller than $J_1$ and $J_2$,
and the Cu(2) ions are mainly coupled via $J_2^b$,
through a Cu(1) in the kagome plane. This is consistent with the magnetic structure in Cu4 \cite{FengZL18}. It suggests that the couplings between the interlayer spins and kagome spins are one order smaller than those within the kagome planes. We expect that the effects of the former should be reduced in Cu3Zn with just 10\% interlayer spins in an undistorted kagome lattice. In addition, the residual interlayer Cu$^{2+}$ ions may not distribute randomly and could form small clusters \cite{FengZL18}, whose spin system may be further isolated from the kagome one.

\begin{figure}[tbp]
\includegraphics[width=\columnwidth]{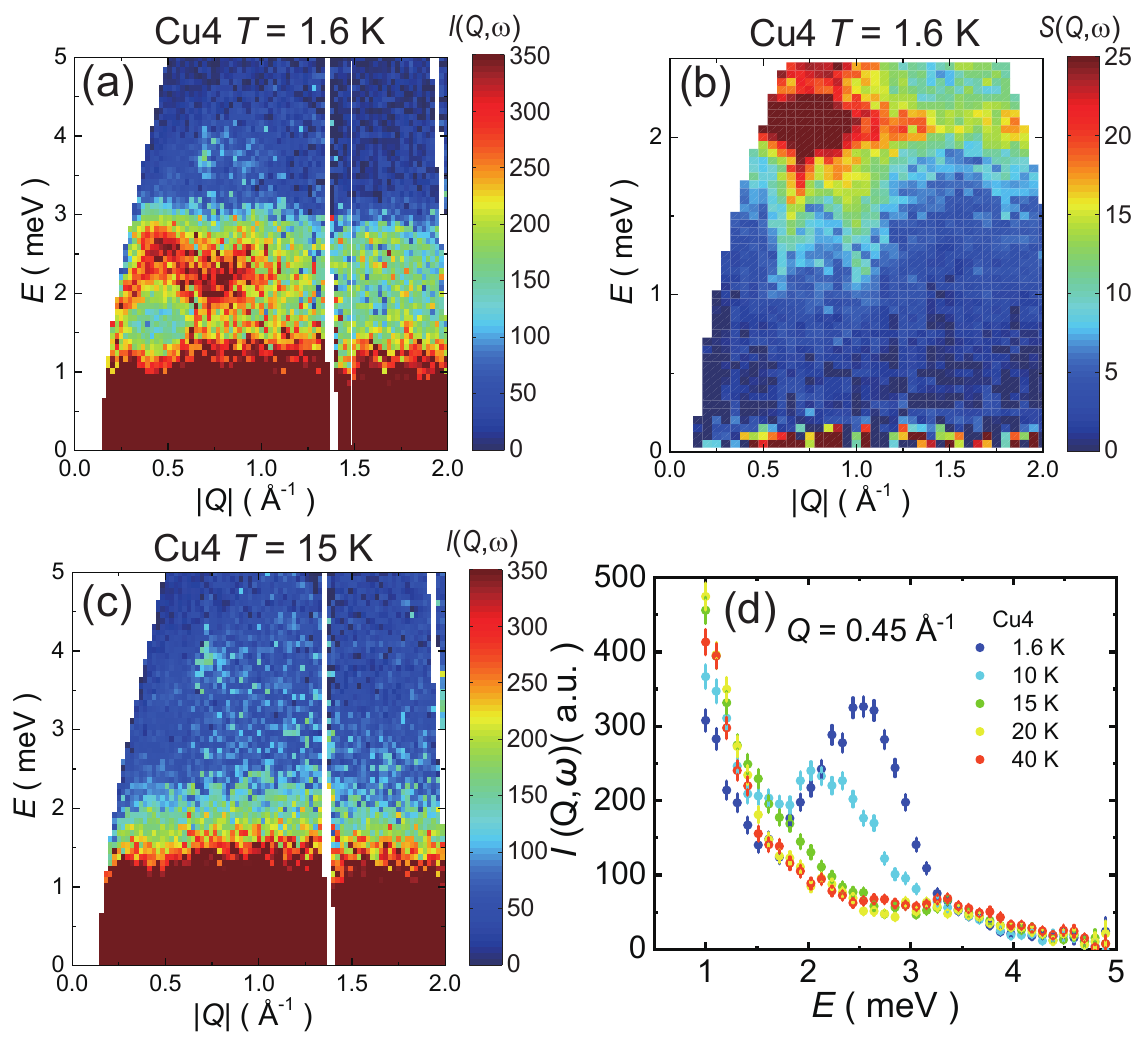}
 \caption{(a) \& (b) Colormaps of the spin-excitations intensities for Cu4 at 1.6 K with $E_i$ = 15 and 3.719 meV, respectively. The unit for the intensities $I(Q,\omega)$ here and below is of arbitrary unit. The unit of $S(Q,\omega)$ is mbarn/meV \AA$^{-1}$ Cu throughout the paper if not mentioned. The spin excitations above 3 meV are centered at about 1.25 \AA$^{-1}$ and too weak to be seen in this scale. (c) Colormaps of the $S(Q,\omega)$ for Cu4 with $E_i$ = 15 meV at 15 K.  (d) Constant-$Q$ cuts at 0.45 \AA$^{-1}$ at different temperatures.}
\label{Cu4:INSLow}
\end{figure}

\subsection{Spin excitations in Cu$_4$(OD)$_6$FBr}

After theoretically demonstrating the couplings between interlayer and kagome spins are weak, we further experimentally show how the kagome and interlayer spin systems in Cu4 are coupled. 
Figures \ref{Cu4:INSLow}(a) and \ref{Cu4:INSLow}(b)  show the low-energy spin excitations of Cu4 at 1.6 K. The band top is below 3 meV, which is consistent with the values of couplings $J_2$ between the interlayer and kagome spins as calculated in previous subsection. Two dispersions emerge at $Q \approx$ 0.65 and 1 \AA$^{-1}$, which correspond to the AFM arrangements of interlayer spins vertical to and within the kagome planes (note that the actual lattice is orthorhombic) \cite{FengZL18,TustainK18}. At $T_N \sim$ 15 K, these features totally disappear as shown in Fig. \ref{Cu4:INSLow}(c). 
Figure \ref{Cu4:INSLow}(d) shows the constant-$Q$ cuts of 0.45 \AA$^{-1}$ at different temperatures. The peak centered at 2.5 meV moves to lower energy at 10 K and completely disappear at 15 K. Above $T_N$, the low-energy spin excitations can be described by a Lorentzian function, suggesting weak spin correlations. 

\begin{figure}[tbp]
\includegraphics[width=\columnwidth]{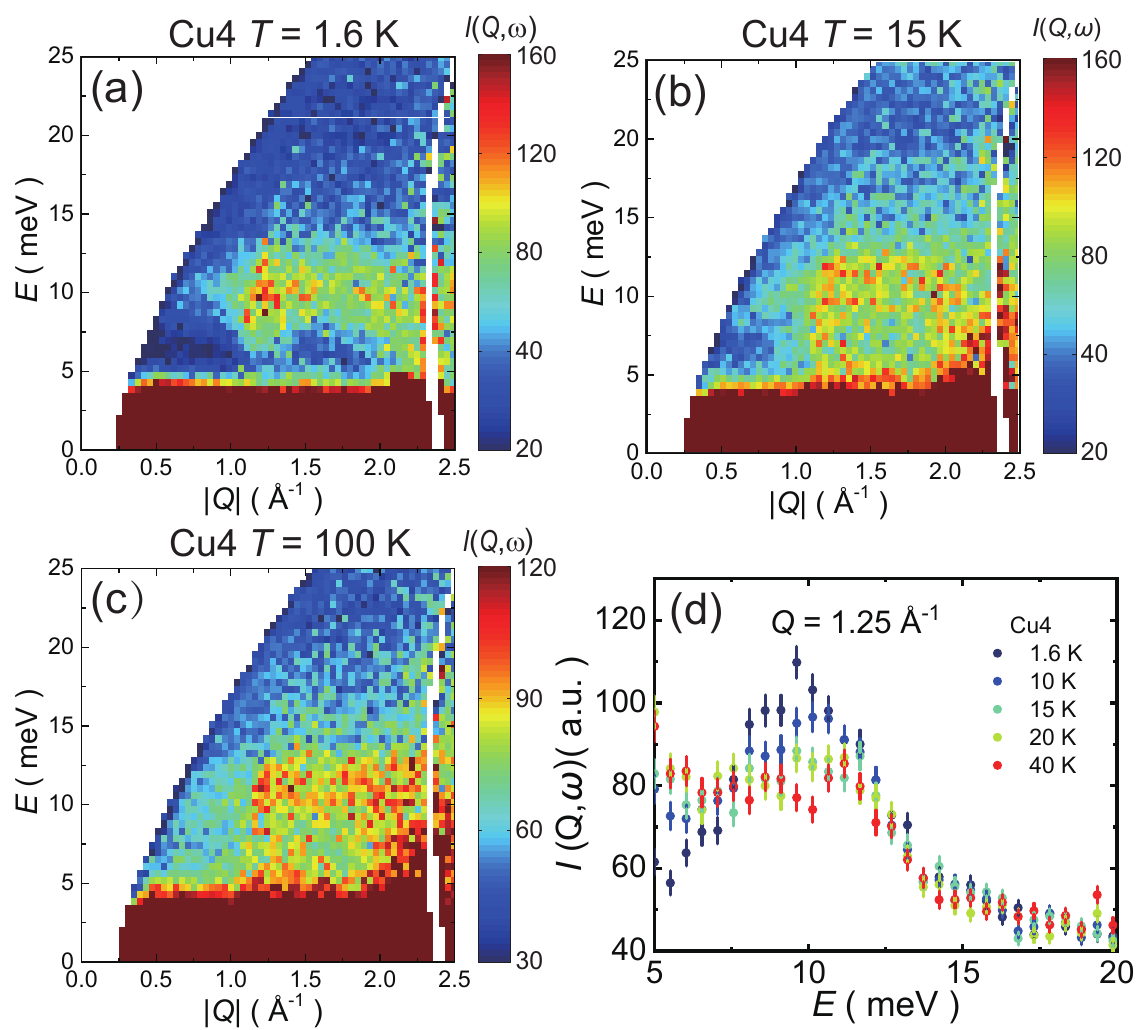}
 \caption{(a) , (b) \& (c) Colormaps of the spin-excitations intensities for Cu4 at 1.6, 15 and 100 K, respectively. The incident neutron energy $E_i$ is 45 meV. The intensities below about 4 meV are too strong to be shown. (d) Constant-$Q$ cuts at 1.25 \AA$^{-1}$ at different temperatures. (e) \& (f) Colormaps of the spin-excitations intensities for Cu4 with $E_i$ = 15 meV at 1.6 and 15 K, respectively. The spin excitations above 3 meV are centered at about 1.25 \AA$^{-1}$ and too weak to be seen in this scale. (g) Colormaps of the $S(Q,\omega)$ for Cu4 with $E_i$ = 3.719 meV at 1.6 K. The unit of $S(Q,\omega)$ is mbarn/meV \AA$^{-1}$ Cu throughout the paper if not mentioned. (h) Constant-$Q$ cuts at 0.45 \AA$^{-1}$ at different temperatures.}
\label{Cu4:INSHigh}
\end{figure}

There are also spin excitations above 3 meV, whose intensities are too weak to be seen in Fig. \ref{Cu4:INSLow}. Figure \ref{Cu4:INSHigh}(a) shows the corlormaps of the INS intensity $I(Q,\omega)$ for Cu4 at 1.6 K. The upper bound for the spin excitations is above 15 meV, which is consistent with the value of $J_1 >$ 10 meV as calculated in previous subsection. It is interesting to note that the excitations seems to show clear boundaries, which need to be further studied on single crystals. The high-energy spin excitations are centered at about 1.25 \AA$^{-1}$, which is very different from those of low-energy spin excitations. While the sample studied here is polycrystalline, we can compare our results with those in herbertsmithite to understand the origin of these high-energy spin excitations. For single-crystal herbertsmithite, INS experiments have revealed that the spin excitations above 1 meV show ring-like excitations within the ($H$,$K$,0) plane with the maximum intensities at (1,0,0) or (0,1,0) \cite{HanTH12,HanTH16}. Vertical to the plane, the spin excitations become maximum around $L$ = 1 \cite{HanTH12,HanTH16}. The powder-averaged high-energy spin excitations for polycrystalline herbertsmithite are thus centered around $|$(1,0,1)$|$ \cite{VriesMA09}. This is the same for Cu4, where the center of high-energy spin excitations is close to the absolute value of $Q$ = (1,0,1) in the hexagonal notation (Here the lattice distortion at low temperatures is neglected). 

With increasing temperature above $T_N$ and up to 100 K (Figs. \ref{Cu4:INSHigh}(b) and \ref{Cu4:INSHigh}(c)), the main features of high-energy spin excitations are still present but become blurred. Figure \ref{Cu4:INSHigh}(d) gives the constant-$Q$ cuts at 1.25 \AA$^{-1}$ for several temperatures. At base temperature, the intensities show a broad peak at about 10 meV. With increasing temperature, the peak intensity decreases and the intensities seem to shift to lower energies. Above $T_N$, the intensities show little change with the temperature. 

To understand the spin excitations in Cu4, we recall that as shown by the theoretical calculations in previous subsection, the dominate couplings in Cu4 are $J_1$'s between the kagome spins, whereas the largest coupling associated with the interlayer spins is $J_2^b$. Compared with the INS data, it is clear that the spin excitations below 3 meV are three-dimensional (3D) spin waves from the long-range magnetic order, which are mainly associated with the interlayer spins. Previous neutron-diffraction results have also shown that the order moments in barlowite mainly come from the interlayer Cu$^{2+}$ \cite{FengZL18,TustainK18}. On the other hand, the spin excitations above 3 meV in Cu4 are from the kagome spins, which are partially affected by the magnetic ordering. 
When the magnetic order disappears above $T_N$, the excitations from the interlayer spins becomes diffusive and are essentially decoupled from those from the kagome spins. This serves as the important starting point to understand the spin excitations in Cu3Zn.

\begin{figure}[tbp]
\includegraphics[width = \columnwidth]{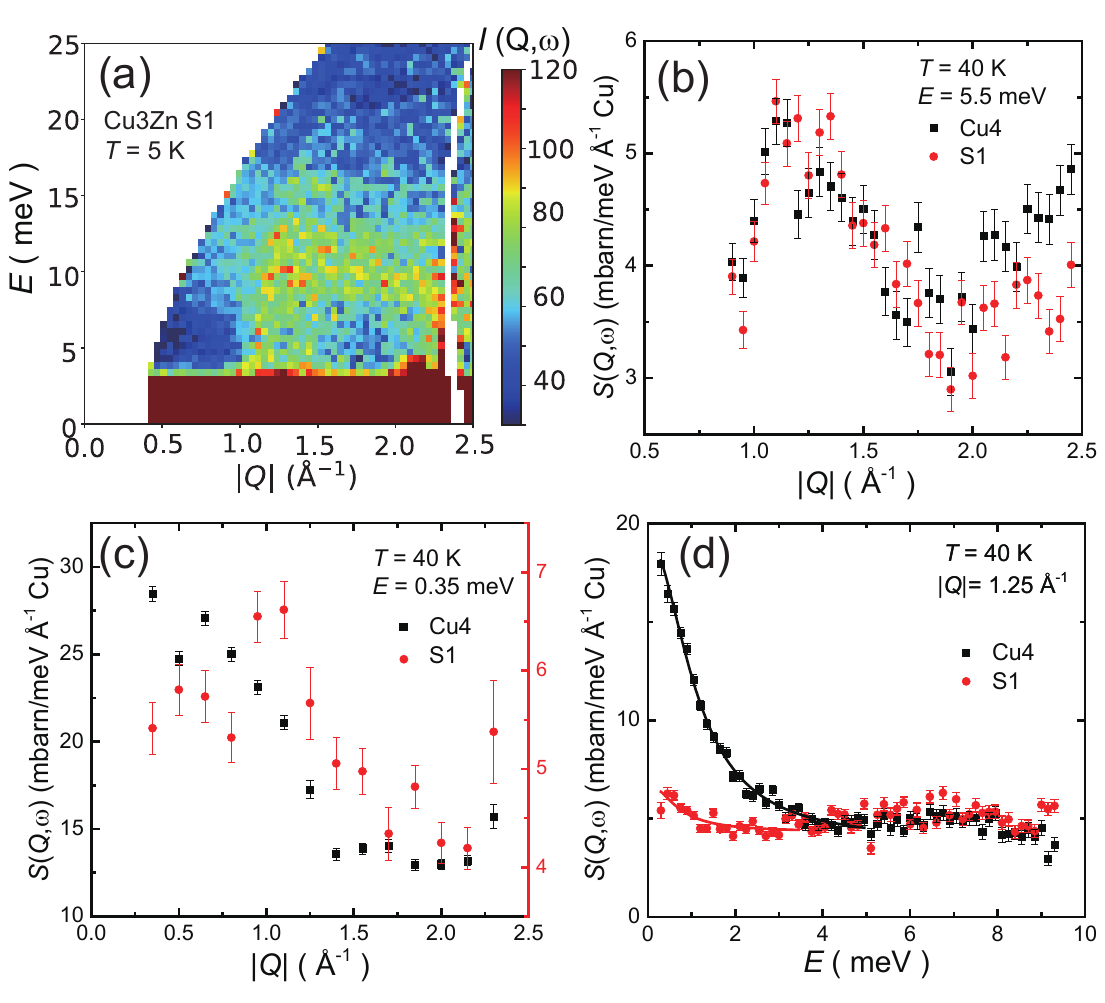}
 \caption{(a) Colormap of $S(Q,\omega)$ for Cu3Zn S1 at 5 K with $E_i$ = 45 meV.  (b) \& (c) Comparison for the Constant-$E$ cuts at 40 K between Cu4 and Cu3Zn S1 at 5.5 and 0.35 meV, respectively. (d) Comparison for the Constant-$Q$ cuts at 40 K between Cu4 and Cu3Zn S1 at 1.25 \AA$^{-1}$. The solid lines are fitted by the Lorentzian function. }
\label{Cu3Zn:INSHigh}
\end{figure}

\subsection{The influence of magnetic impurities on spin excitations of Cu$_3$Zn(OH)$_6$FBr}

Figure \ref{Cu3Zn:INSHigh}(a) shows the spin excitations of Cu3Zn S1 at 5 K, which are very similar to those of Cu4 above $T_N$ (Figs. \ref{Cu4:INSHigh}(b)) \& \ref{Cu4:INSHigh}(c)), including similar band tops and the overall features. In fact, we find that the spin excitaions between Cu4 and Cu3Zn can be quantitatively compared. Figure \ref{Cu3Zn:INSHigh}(b) shows the constant-$E$ cuts at 5.5 meV for Cu4 and Cu3Zn S1 at 40 K, which are quantitatively very close to each other. At low energies, the $Q$ dependence of these two samples are also similar, as shown in Fig. \ref{Cu3Zn:INSHigh}(c). However, the intensities are very different. The constant-$Q$ cuts at 1.25 \AA$^{-1}$ (Fig. \ref{Cu3Zn:INSHigh}(d)) further reveal that the high-energy spin excitations of Cu4 and Cu3Zn S1 are almost the same but the low-energy ones are different. Fitting the low-energy $S(Q,\omega)$ with a Lorentzian function gives the ratio between the areas of CuZn S1 and Cu4 as about 0.1, which is close to the concentration of the interlayer Cu$^{2+}$ ions in the former. These results suggest that the low-energy spin excitations of Cu3Zn S1 at 40 K are mainly from the residual interplayer spins.

\begin{figure}[tbp]
\includegraphics[scale = 0.45]{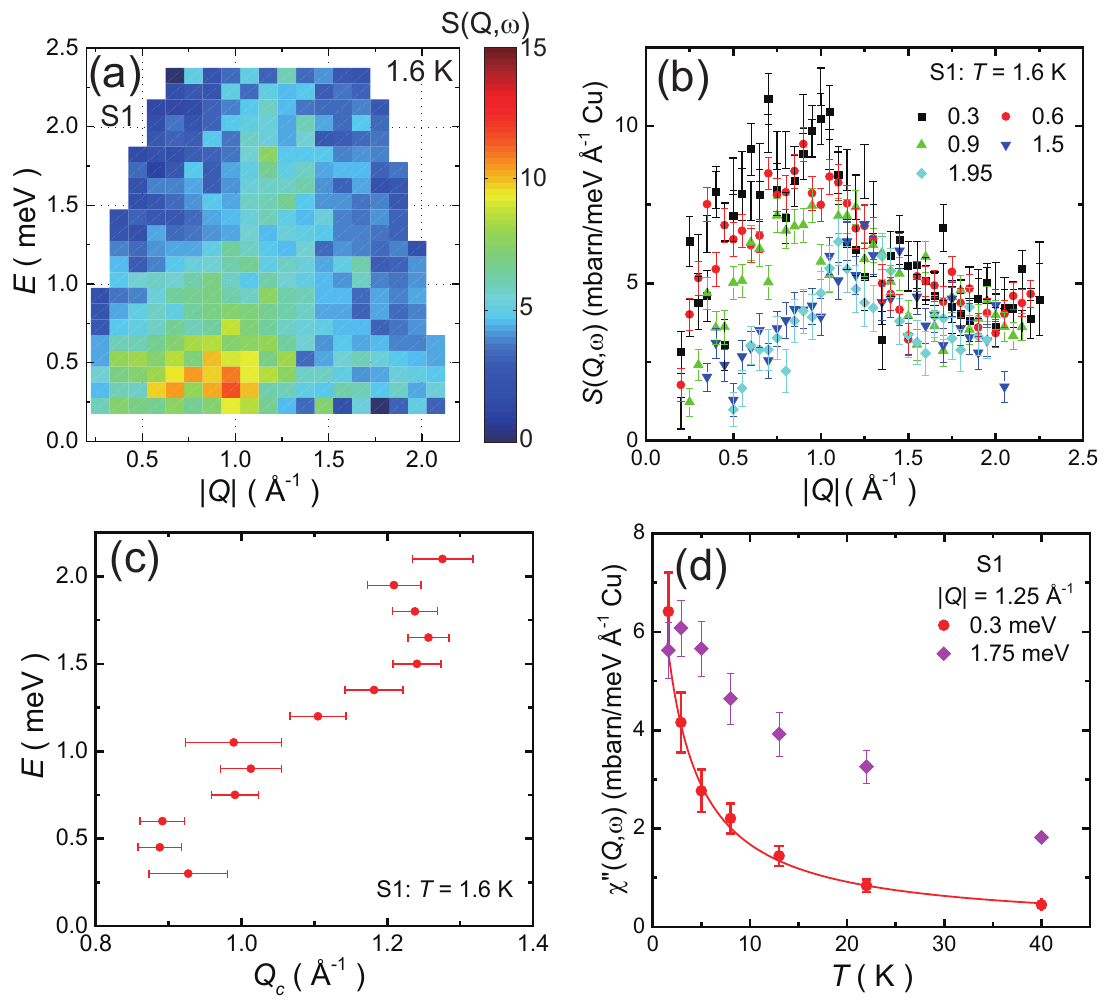}
 \caption{(a) Colormap of the $S(Q,\omega)$ for Cu3Zn S1 at 1.6 K with $E_i$ = 3.719 meV. (b) Constant-$E$ cuts from the data in (a). (c) The change of central momentum value $Q_c$ with $E$. Here $Q_c$ is determined from the fitting of the constant-$E$ cuts with a Lorentzian function as described in the main text. (d) Temperature dependence of $\chi"(Q,\omega)$ at $|Q|$ = 1.25 \AA$^{-1}$ for $E$ = 0.3 and 1.75 meV. The solid line is fitted by the Curie-Weiss function as described in the main text.}
\label{Cu3Zn:INSLow}
\end{figure}

The results at low temperatures reveal that the influence of magnetic impurities is still concentrated at low energies. Figure \ref{Cu3Zn:INSLow}(a) shows the colormap of low-energy spin excitations of Cu3Zn S1 at 1.6 K. The constant-$E$ cuts at different energies are shown in Fig. \ref{Cu3Zn:INSLow}(b). The $Q$ dependence of $S(Q,\omega)$ can be roughly fitted by a Lorentzian function, which should not precisely describe the $|Q|$ dependence of the spin excitations but is good enough to give a sense of the change of the spectra weight with energy. As shown in Fig. \ref{Cu3Zn:INSLow}(c), the fitted center $Q_c$ only changes between about 0.7 to 1.5 meV. For $E >$ 1.5 meV, the spin excitations are centered at 1.25 \AA$^{-1}$ as the high-energy spin excitations at 40 K. Below 1 meV, the spectra weight shift to lower $Q$, also similar to that at 40 K. This can be simply understood as the spin excitations centered at 0.9 \AA$^{-1}$ become weaker with increasing energy, rather than some kind of intrinsic dispersion-like behavior for the kagome spin system. To see whether the high-energy spin excitations persist at lower energies, we check the temperature dependence of the imaginary part of the dynamical susceptibility $\chi"(Q,\omega)$ = $S(Q,\omega)$(1-exp(-$E$/k$_BT$)). At 0.3 meV and 1.25 \AA$^{-1}$, $\chi"(Q,\omega)$ can be fitted by the Curie-Weiss function $A/(T-\theta)$ with the value of $\theta$ as -1.44 K (Fig. \ref{Cu3Zn:INSLow}(d)), which is very close to -1.28 K obtained from fitting the DC magnetic susceptibility at low temperatures \cite{supp}. On the other hand, the $\chi"(Q,\omega)$ at 1.75 meV decreases with temperature at a much slower rate. These results clearly demonstrate that the spin excitations from the kagome and interlayer spin systems are well separated and the latter dominates the low-energy spin excitations in Cu3Zn. This is also consistent with our theoretical calculations given above, which show that the couplings between the interlayer Cu$^{2+}$ spins and those within the kagome planes are indeed weak. 

\subsection{Spin gap in Cu$_3$Zn(OD)$_6$FBr}

\begin{figure}[tbp]
\includegraphics[scale=0.45]{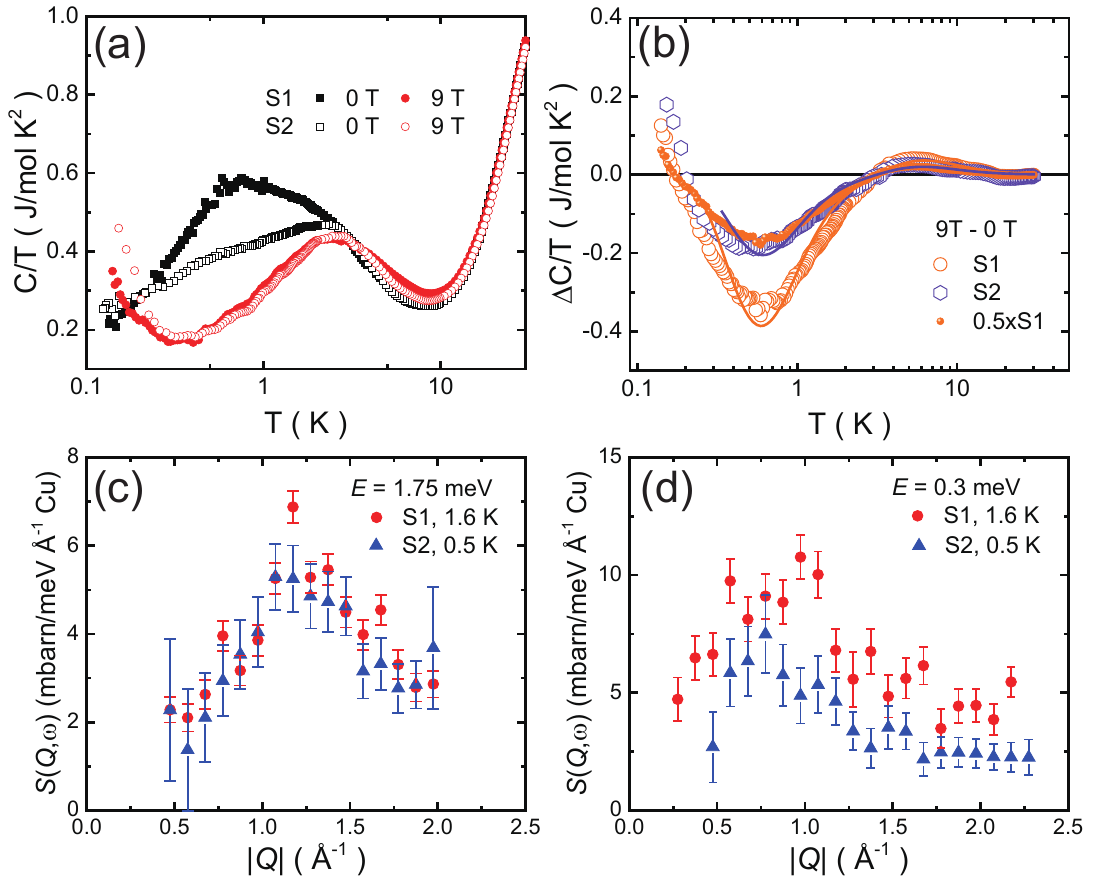}
\caption{(a) $C/T$ at 0 and 9 T for Cu3Zn S1 and S2. (b) The difference of the specific heats between 9 and 0 T. The solid lines are fitted by the Schottky anomaly function for magnetic impurities as described in the main text. The low-temperature upturns are most likely due to nuclear Schottky anomalies, which are more prominent in hydrogen samples \cite{FengZL17}. The solid red dots are $\Delta C/T$ of S1 divided by 2. (c) \& (d) Constant-$E$ cuts for Cu3Zn S1 and S2 at 1.75 and 0.3 meV, respectively.}
\label{Cu3ZnS1S2}
\end{figure}

Based on the awareness that the low-energy spin excitations in Cu$_3$Zn(OD)6FBr mainly come from the magnetic impurities, our protocol continues to demonstrate how to reveal the intrinsic spin excitations within the kagome planes from experimental data and as shown in this and next subsections, that they are gapped in Cu3Zn. To this end, we take a route of comparing the spin excitations of two Cu3Zn samples, S1 and S2, which have different Zn contents because of different synthesizing temperatures.

As shown in herbertsmithite \cite{VriesMA08}, the magnetic impurities mainly contribute to the specific heat at low temperatures and can be easily affected by the magnetic field due to their weak correlations. Figure \ref{Cu3ZnS1S2}(a) shows the specific heat for both Cu3Zn S1 and S2. At 0 T, the values of $C/T$ for them are largely different below about 2.5 K, but become almost the same above, suggesting the contributions from magnetic impurities at low temperatures due to their different concentration in Cu3Zn S1 and S2. At 9 T, the low-temperature specific heats become almost the same because those from magnetic impurities are pushed to higher temperatures. Similar to herbertsmithite \cite{VriesMA08}, the amount of magnetic impurities $N_m$ can be estimated from fitting the difference of the specific heat between 9 and 0 T, i.e. $\Delta C$, by the Schottky-anomaly function as shown in Fig. \ref{Cu3ZnS1S2}(b). The results give 7.2\% and 4\% of $N_m$ for S1 and S2, respectively. The fittings at low temperatures are not good, which suggests the presence of spin correlations even for diluted magnetic impurities \cite{HanTH16}. By simply plotting $\Delta C/T$ for S2 and half of that for S1 together as shown in Fig. \ref{Cu3ZnS1S2}(b), it can be seen that the major difference of the specific heats between S1 and S2 comes from their difference in $N_m$ and $N_m$ of S1 is about twice of that of S2.

Since the spin excitations within the kagome planes are very weakly coupled to those from magnetic impurities as shown in the previous subsection, the INS data of these two samples should be also able to reveal this information. Figures \ref{Cu3ZnS1S2}(c) and \ref{Cu3ZnS1S2}(d) show the constant-$E$ cuts of spin excitations at 1.75 and 0.3 meV, respectively. There is no difference for S1 and S2 in the former, whereas the $S(Q,\omega)$ ($Q >$ 1 \AA$^{-1}$) for S1 is much larger than that for S2 at 0.3 meV. This is consistent with the results in previous subsection that the low-energy spin excitations are dominated by magnetic impurities. 

Accordingly, the spin excitations from magnetic impurities, $S_{imp}(Q,\omega)$, can be simply removed by the following method. The total intensity $S_{tot}(Q,\omega)$ is comprised of $S_{kag}(Q,\omega)$ from kagome planes and $S_{imp}(Q,\omega)$ as
\begin{equation}
S_{tot}(Q,\omega) = S_{kag}(Q,\omega) + S_{imp}(Q,\omega).
\end{equation}
As shown above, $S_{kag}(Q,\omega)$ is almost the same for CuZn S1 and S2, while $S^{S1}_{imp}(Q,\omega) \approx 2S^{S2}_{imp}(Q,\omega)$, therefore one can easily see
\begin{eqnarray}
S^{S2}_{tot}(Q,\omega)-S^{S1}_{tot}(Q,\omega) &\approx& S^{S2}_{imp}(Q,\omega) - S^{S1}_{imp}(Q,\omega) \nonumber\\
 &\approx& -S^{S2}_{imp}(Q,\omega).
\end{eqnarray}
This means that
\begin{eqnarray}
S_{kag}(Q,\omega) &=& S^{S2}_{tot}(Q,\omega) - S^{S2}_{imp}(Q,\omega) \nonumber\\
 &=& 2S^{S2}_{tot}(Q,\omega)-S^{S1}_{tot}(Q,\omega),
\end{eqnarray} 
which suggests that due to the twice difference in the amount of impurities in Cu3Zn S1 and S2, one can directly calculate the spin excitations from the kagome planes without any particular models for the spin correlations of magnetic impurities .

\begin{figure}[tbp]
\includegraphics[scale=0.45]{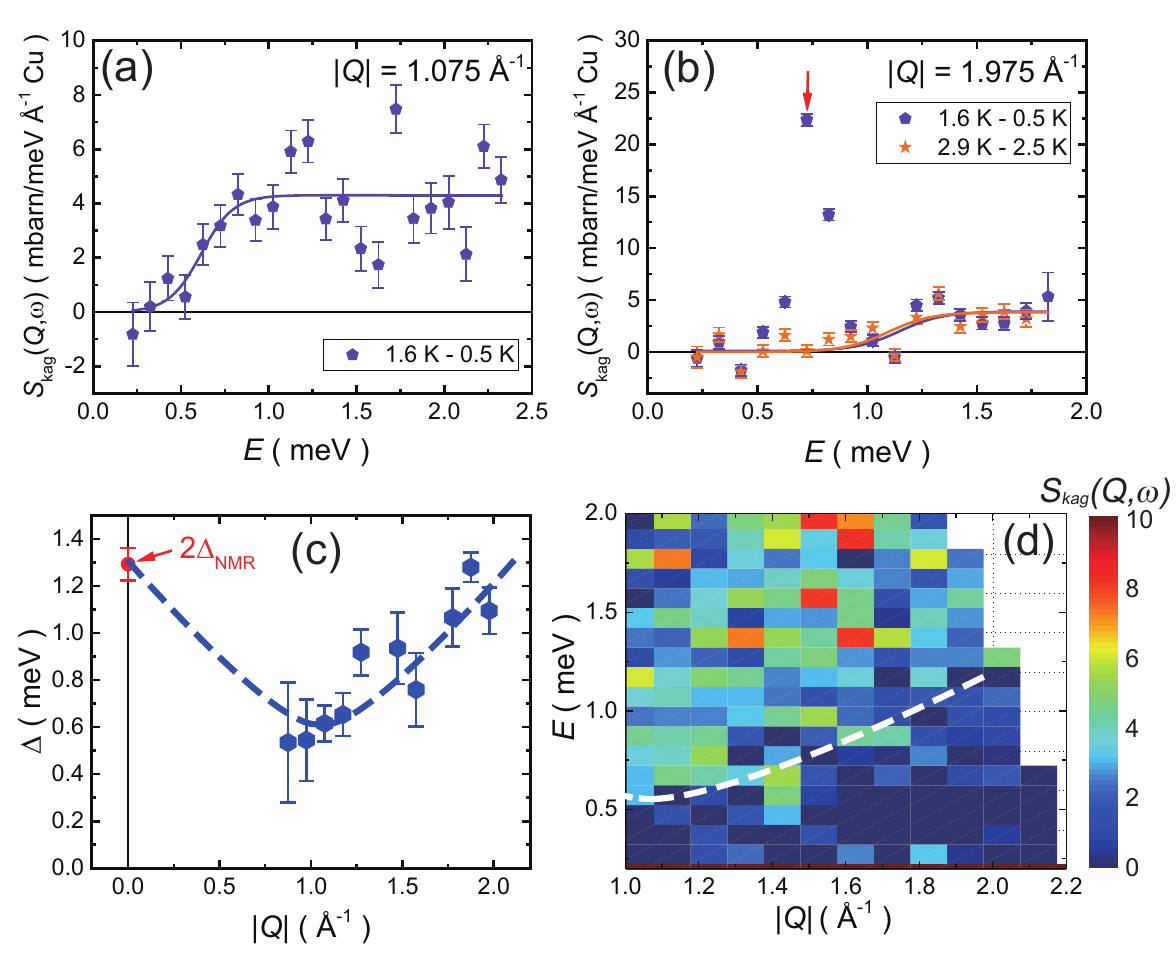}
\caption{(a) \& (b) Constant-$Q$ cuts for the spin excitations of the kagome planes at $|Q|$ = 1.075 and 1.975 \AA$^{-1}$, respectively. The solid lines are fitted results by the gap function as described in the main text. The first and second temperatures in the labels are the temperatures for S1 and S2, respectively. The sharp peak around 0.75 meV in (b) is from the $^4$He roton mode from the liquified Helium in the sample can, which quickly disappears at higher temperatures.
(c) $|Q|$ dependence of spin gap in Cu3Zn. The data at $|Q|$ = 0 is 2$\Delta_{NMR}$, where $\Delta_{NMR}$ is the spinon gap value obtained from the NMR measurements \cite{FengZL17}. The dashed line is guided to the eyes with minimum value at $|Q|$ = 1.088 \AA$^{-1}$, which corresponds to (1,0,0). The error bars are from the fitting by the gap function as described in the main text. (d) Colormap of the $S(Q,\omega)$ for the kagome planes in Cu3Zn. To improve the statistics, data below 5 K are summed up. The dashed line is guided to the eyes.}
\label{Cu3Zn:Gap}
\end{figure}

The results of $S_{kag}(Q,\omega)$ at two different $|Q|$'s are shown in Figs. \ref{Cu3Zn:Gap}(a) and \ref{Cu3Zn:Gap}(b), which shows that $S_{kag}(Q,\omega)$ at low energies is essentially zero. The data can be fitted by a gap function $A(1+\tanh(\frac{E-\Delta}{\Gamma}))$ with $\Gamma$ fixed to 0.2 meV \cite{HanTH16}, where $A$, $\Delta$ and $\Gamma$ are fitting parameters. Figure \ref{Cu3Zn:Gap}(c) shows the $|Q|$ dependence of the gap value $\Delta$ in Cu3Zn \cite{supp}. Figure \ref{Cu3Zn:Gap}(d) gives the colormap of the spin excitations from the kagome planes, which further demonstrates the gapped excitations.

\subsection{Possible vison excitations in Cu$_3$Zn(OD)$_6$FBr}

\begin{figure}[tbp]
\includegraphics[scale=0.45]{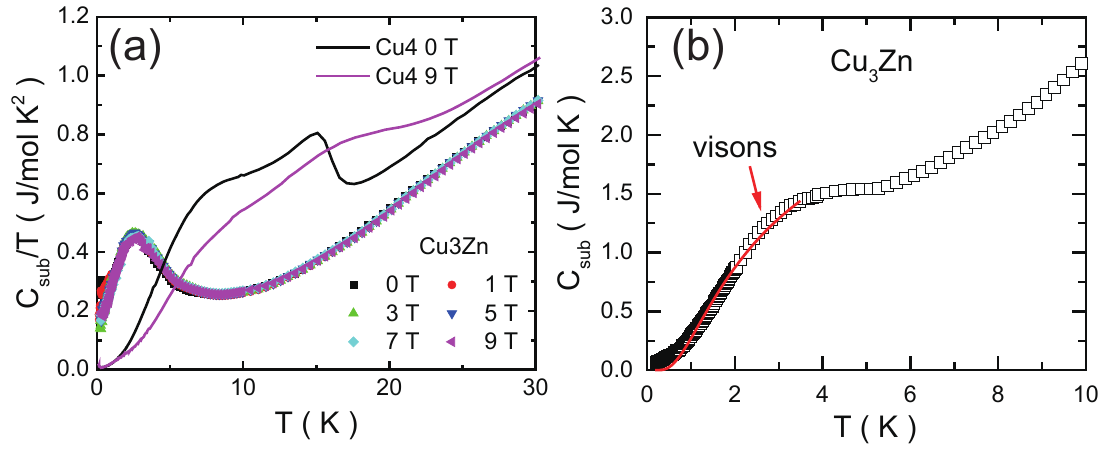}
\caption{ The specific heat of Cu3Zn after subtracting the contribution from magnetic impurities at different fields. The solid lines are $C/T$ of Cu4 at 0 and 9 T. (b) The subtracted specific heat $C_{sub}$ of Cu3Zn at 0 T. We note that there is essentially no difference for fields from 0 to 9 T. The solid line is fitted by the gap function as described in the main text.}
\label{HC}
\end{figure}

Since the spin excitations of kagome planes are gapped, one may expect that the specific heat from the kagome spin system, $C_{kag}$, should also show gap behavior. Similar to the process in dealing with the INS data, $C_{kag}$ can also be derived by subtracting the contribution from magnetic impurities, $C_{imp}$. The total specific heat is 
\begin{equation}
C_{tot} = C_{ph} + C_{kag} + C_{imp},
\end{equation} 
where $C_{ph}$ is the phonon specific heat. Since $C_{ph}$ and $C_{kag}$ should change little with the concentrations of magnetic impurities and $C^{S1}_{imp} \approx 2C^{S2}_{imp}$, we have $C^{S2}_{imp} \approx C^{S1}_{tot}-C^{S2}_{tot}$. Figure \ref{HC}(a) shows $C_{sub}$ = $C^{S2}_{tot}$ - $C^{S2}_{imp}$ = $C_{kag}$ + $C_{ph}$. Because the values of $C_{ph}$ are very small below 5 K as indicated from the specific heat of Cu4, it is clear that $C_{kag}$ below 5 K shows almost no magnetic field dependence. The field-independent behavior of magnetic specific heat has also been found in the kagome-like magnet SrCr$_{9p}$Ga$_{12-9p}$O$_{19}$ \cite{RamirezAP00}, which has been attributed to the many-body singlet excitations with $S$ = 0. These $S$ = 0 excitations are visons in $Z_2$ QSLs on the kagome lattice \cite{NikolicP03,SchnackJ18,SunGY18}. Moreover, since the vison excitations should also be gapped in a $Z_2$ topological order, the gap value can be estimated by fitting the low temperature data by exp(-$\Delta_v$/T) as shown in Fig. \ref{HC}(b), which gives $\Delta_v \sim$ 2.3 K, or 0.2 meV, much lower than the lowest singlet-triplet gap $\Delta \sim 0.6$ meV as shown in the above subsection. These results are consistent with theoretical calculations for the specific heat of a two-dimensional kagome lattice \cite{SchnackJ18}. 

\section{discussions}

The experimental results in this work can be summarized as follows: (i) The spin excitations in Cu3Zn can be well separated into two parts, with high-energy and low-energy spin excitations corresponding to the kagome spins and interlayer spin impurities, respectively; (ii) The intrinsic kagome spin excitations of Cu3Zn are gapped and show continuum above the gap; (iii) The intrinsic kagome magnetic specific heat of Cu3Zn at low temperatures is immune to magnetic field. It should be pointed out that the first-principle calculations of the magnetic couplings in Cu3Zn and Cu4 are consistent with result (i). It is the result (i) that enables us to reliably analyze the INS and specific heat data to obtain the information of excitations within the kagome planes, and from (ii) and (iii) we then successfully reveal the two anyonic excitations -- the hallmark of the $Z_2$ QSL -- from the INS and specific-heat results. Our results and the data analysis protocol therefore provide the solution to the long-standing problem from the early days of herbertsmithite and other similar systems, i.e., how to extract the intrinsic QSL features from such system in the presence of the residual moments outside of the kagome planes.

While the spin continuum has always treated as a solid, sometimes the best, evidence for the existence of spinons, it is actually just a necessary, not a sufficient condition. In other words, some other ground states or the presence of strong disorders may also result in low-energy spin continuum~\cite{JSWen2018,Kimchi2018}. Result (ii), combined with our previous NMR result, may actually provide a sufficient condition due to the unique property of the $Z_2$ topological order, i.e., all excitations are gapped. It has been suggested in the NMR experiment that a spin gap $\Delta_{NMR}$ with $S$ = 1/2 exists in Cu3Zn \cite{FengZL17}, one would expect that the gap value $\Delta$ from the INS experiment should be about twice of $\Delta_{NMR}$ because neutron only detects two spinons with $S$ = 1. We note that this comparison should be done at $Q$ = 0 since the Knight shift only measuring the uniform magnetic susceptibility. While it is impossible to measure $Q$ = 0 spin excitations directly in the INS experiment, the periodic nature of the spin excitations enable us to obtain $\Delta$ at $Q$ = 0. As shown in herbertsmithite, the periodicity of spin excitations within the ($H$,$K$,0) plane is twice of that of the kagome lattice in the reciprocal space \cite{HanTH12,HanTH16}, which is also consistent with our results. Accordingly, the spin gap should also follow this periodicity as shown by the dashed line in Fig. \ref{Cu3Zn:Gap}(c), which shows that $\Delta$ at $Q$ = 0 is indeed 2$\Delta_{NMR}$. Therefore, the spin excitations and gap measured in the INS experiment here are from pairs of spinons.

Result (iii), i.e., the large field-independent specific heat at low temperatures, helps us to identify the vison excitations in the QSLs for the first time. As suggested above, the vison excitations are gapped with a gap value of about 0.2 meV, which is much smaller than the lowest singlet-triplet gap value of spinons ($\sim$ 0.6 meV). The entropy release in the kagome-lattice quantum dimer model, which comes solely from visons, is one-third of Rln2~\cite{FaWang2007}. It should be pointed out that this model overestimates the vison entropy in a spin model because the spin-singlet pairs are not orthogonal, unlike the dimers in the quantum dimer model. On the other hand, numerical simulations have shown that the singlet-singlet excitations have an entropy close to 0.2 Rln2 \cite{SchnackJ18}. In our case, the magnetic entropy below 3 K is about 0.2 Rln2, which is a reasonable value for visons. We note that the spinon gap should close at about 10 T \cite{FengZL17} but here no indication can be seen from the specific heat data, which indicates that the specific heat induced by closing the gap is small \cite{supp}. One possible scenario is that the gap is closed at some particular $Q$'s rather than uniformly so the field-induced spinon's densities of states are low. 

While the above discussions take a very optimistic view for the existence of $Z_2$ topological order in Cu$_3$Zn(OH)$_6$FBr, there could be several issues that undermine the conclusions. First, the analysis to derive the spin gap cannot be done for $|Q| <$ 0.9 \AA$^{-1}$ (Figs. \ref{Cu3Zn:Gap}(c) and \ref{Cu3Zn:Gap}(d))because the low-energy spin excitations do have a simple 2:1 ratio for Cu3Zn S1 and S2, as shown in Fig. \ref{Cu3ZnS1S2}(d). This is most likely because the residual interlayer Cu$^{2+}$ ions tend to form small clusters \cite{FengZL18,WeiYuan20Magnetic}, which should result in short-range spin-spin correlations that are Zn-concentration dependent. However, the larger $|Q|$ data still provide good results and because the periodicity of the spin excitations, the $|Q|$ dependence of the gap in Fig. \ref{Cu3Zn:Gap}(c) is still reliable. Second, the gap for the specific-heat data may be too optimistic as one cannot actually observe the nearly zero value of the specific heat at very low temperatures for the gapped excitations due to the existence of nuclear Schottky anomalies. However, we would like to point it out that our recent studies on the size-dependent specific-heat studies also support the existence of gapped excitations \cite{WeiYuan20Nonlocal}. Moreover, the analysis provides consistent understandings on the possible $Z_2$ toplogical order in Cu$_3$Zn(OH)$_6$FBr and gives a natural explanation on why no gap behavior is found in the specific heat. Third, it is rather surprising to see that high-energy spin excitations in Cu4 above $T_N$ are very similar to those in Cu3Zn, which needs to be further studied to see whether the spin liquid state already leave its trace even in Cu4 as indicated previously from the magnetic-susceptibility measurements \cite{HanTH16}. Despite the above issues, we emphasize that it is the consistency of all the experimental and theoretical analysis that helps us identify the ground state of Cu3Zn. 

Our results provide strong evidences for the existence of gapped spin excitations and gapped singlet-singlet excitations in Cu3Zn, which correspond to the spinons and visons in a $Z_2$ topological order. The material realization of such state provides us opportunities to further investigate the unique properties of $Z_2$ topological orders by directly comparing theoretical and experimental results in the future. For example, it has been shown theoretically that a topological order is robust against local perturbations \cite{HastingsMB05}. If the ICP results can be trusted, the Cu3Zn S2 sample may have about 5\% Zn within the kagome planes, which seems to have negligible effect on either the spin dynamics nor the specific heat of the kagome planes. Another important question is how the spinons and visons are coupled. Theoretical simulation has shown that strong coupling between spinons and visons can alter the spin excitations dramatically \cite{PunkM14,SunGY18}. While our samples are polycrystalline, the $|Q|$ dependence of the spin excitations still suggests that the gap mainly reflect the in-plane average result so that there may have very low-energy spin excitations at some particular points near zone boundaries (e.g. $Q$ = (2,0,0) or (1, 1, 0)), which should come from the bound states between spinons and visons. Whether this is true or not of course needs to be further studied both experimentally and theoretically but at least the momentum dependence in INS spin spectra is consistent with picture of spinon-vison coupling. Last but not least, it is highly desirable to see whether the $Z_2$ topological order in Cu3Zn can be tuned into a symmetry-breaking order or trivial paramagnetism by a quantum critical transition, such a critical point between topological order and symmetry-breaking is also at the theoretical and experimental frontiers in understanding the new paradigms of quantum material beyond conventional wisdom.
	
\section{conclusions}
We show that the ground state of Cu$_3$Zn(OH)$_6$FBr may be a $Z_2$ topological order by showing evidences for its two gapped fractionalized excitations, spinons and visons, which were detected by the inelastic neutron scattering and specific heat techniques, respectively. We have
also performed first principle calculations to estimate the coupling 
strengths of both the Cu4 and Cu3Zn compounds, which are 
consistent with the interpretation of experimental observations. The spinons result in continuum spin excitations as seen in many other QSL candidates, and also the spin gap due to the unique properties of the topological orders. Moreover, the gap value of spin excitations ( $S$ = 1 ) at $Q$ = 0 is about twice of the spinon gap ( $S$ = 1/2 ) measured in the NMR experiments \cite{FengZL17}, which further confirms the nature of the spin excitations. Below the singlet-triplet gap, gapped singlet-singlet excitations are also found, which corresponds to visons. While our results provide many consistent evidences for the existence of a $Z_2$ topological order in Zn-doped Barlowite, we believe this is the beginning of more systematic investigations, both experimental and theoretical, for example on the exact ground state of quantun Kagome Heisenberg antiferromagnetic model and the single-crystal synthesis of the 
compound and its neutron scattering, etc, to eventually pin down the material realization of the topological order with its unconventional 
anyonic excitations in frustrated quantum magnets.

\acknowledgments
We thank Young Lee for insightful discussions on our results in specific and the physics of kagome antiferromagnets in general. This work is supported by the National Key R\&D Program of China (Grants No. 2017YFA0302900, 2016YFA0300502, 2016YFA0300604), the National Natural Science Foundation of China (Grants No. 11874401, 11674406, 11574359, 11674370, 11774399, 11474330), the Strategic Priority Research Program of the Chinese Academy of Sciences (Grants No. XDB25000000, XDB28000000, XDB07020300, XDB07020200). Research Grants Council of Hong Kong Special Administrative Region of China through Grant No. 17303019. Research conducted at ORNL's High Flux Isotope Reactor was sponsored by the Scientific User Facilities Division, Office of Basic Energy Sciences, US Department of Energy. The access to PELICAN instrument for the INS measurements (P6589) at ANSTO is greatly acknowledged. ZYM thanks the hospitality of the Aspen Center for Physics supported by National Science Foundation grant NSF PHY-1066293, where part of this work was developed.

\end{document}